\documentstyle [12pt,twoside]{article}
\newskip\humongous \humongous=0pt plus 1000pt minus 1000pt
\def\caja{\mathsurround=0pt}

\newif\ifdtup
\def\panorama{\global\dtuptrue \openup2\jot \caja
        \everycr{\noalign{\ifdtup \global\dtupfalse
        \vskip-\lineskiplimit \vskip\normallineskiplimit
        \else \penalty\interdisplaylinepenalty \fi}}}

\def\eqalignnotwo#1{\panorama \tabskip=\humongous
        \halign to\displaywidth{\hfil$\displaystyle{##}$
        \tabskip=0pt&$\displaystyle{{}##}$
        \tabskip=\humongous&\llap{$##$}\tabskip=0pt
        \tabskip=0pt&$\displaystyle{{}##}$\hfil
        \crcr#1\crcr}}

\def\begintitle#1#2#3#4
        {\begin{titlepage}
         \centerline{#1 \hfill UMDGR-93-#2}
         \begin{center}\vglue .4in
         {\large\bf #3}\\[.4in]
         {\bf #4}\\[.2in]
         {\it Department of Physics}\\
         {\it University of Maryland, College Park, MD 20742}\\[.4in]
         {\bf ABSTRACT}\\
         \end{center}
         \begin{quotation}}
\def\endtitle
         {\end{quotation}
          \end{titlepage}
          \newpage}

\def\a{\alpha}\def\d{\delta}\def\e{\epsilon}
\def\n{\eta}\def\l{\lambda}
\def\u{\mu}\def\v{\nu}\def\s{\sigma}\def\t{\tau}
\def\w{\omega}
\def\D{\Delta}\def\S{\Sigma}

\def\pd{\partial}\def\grad{\nabla\!}

\def\ovr{\overline}\def\und{\underline}
\def\otw{\widetilde}
\def\utw#1{\rlap{\lower1ex\hbox{$\sim$}}#1{}}
\def\pb#1{\rlap{\lower1ex\hbox{$\leftarrow$}}#1{}}
\def\pf#1{\rlap{\lower1ex\hbox{$\rightarrow$}}#1{}}
\def\tprod{\otimes}

\def\3#1{{}^3\!#1}\def\4#1{{}^4\!#1}\def\+#1{{}^+\!#1}\def\-#1{{}^-\!#1}
\def\*#1{{}^*\!#1}

\begin{document}

\begintitle{February 1993}{28}{\hfil SCALAR AND SPINOR
FIELDS\hfil\break
IN SIGNATURE CHANGING SPACETIMES}{Joseph
D. Romano\footnote{romano@umdhep.umd.edu} }
The propagation of scalar and spinor fields in a spacetime whose metric changes
signature is analyzed. Recent work of Dray {\it et al.}~on particle production
from signature change for a (massless) scalar field is reviewed, and an attempt
is made to extend their analysis to the case of a (massless) spin-half field.
In contrast to their results for a scalar field, it is shown here---for
$SL(2,C)$ spinors---that although there are inequivalent forms of the Dirac
equation that can be used to propagate a spinor in a signature changing
spacetime, none of these forms gives rise to a conserved inner product on the
space of solutions to the field equations.

\vspace{.7cm}


\endtitle

\noindent{\bf 1. Introduction}
\vskip .5cm

It is by now a well-known result of quantum field theory in curved spacetime
that particles are produced in general by a time-dependent, background
gravitational field \cite{B&D}.  Indeed, as early as 1966, Parker in his Ph.D.
thesis \cite{Parker} showed that particle production occurs in an expanding
Robertson-Walker universe.  Given a spacetime that is asymptotically
Minkowskian at early and late times (so that ``in" and ``out" particle states
can be defined), one can turn the crank, so to speak, and follow a fairly
standard procedure for determining whether or not particle production occurs.
One defines ``in" and ``out" vacuum states in terms of ``in" and ``out"
creation and annihilation operators; then writes down the Bogolubov
transformations relating these operators; and then calculates the expectation
value of the number operator for ``out" particles in the ``in" vacuum
\cite{B&D}.  That's all there is to it.  It's fairly simple in principle,
although in practice one may run into mathematical difficulties trying to solve
the differential equations.

Just recently, Dray, Manogue, and Tucker \cite{DM&T} have considered the
question of particle production in a spacetime whose {\it metric changes
signature}. Although such a spacetime is a rather strange background on which
to do quantum field theory, Dray {\it et al.}~were motivated to look at this
problem to gain new insights into similar issues involving particle production
in spacetimes that have degenerate metrics or topology change
\cite{A&D,MC&D,Horowitz}. For instance, they propose a method for analyzing
particle production in the ``trousers" spacetime \cite{A&D,MC&D} that involves
a smooth embedding of the trousers in 3-dimensional Minkowski spacetime.  The
induced metric on the trousers will have Euclidean signature in a region near
the crotch.   To understand this more complicated problem, Dray {\it et
al.}~began by studying the evolution of a (massless) scalar field through a
Euclidean region in an otherwise flat, Lorentzian spacetime.

In their work \cite{DM&T}, Dray {\it et al.}~showed that particle production
occurs for a scalar field propagating in a signature changing spacetime. They
observed that in such a spacetime there are inequivalent forms of the
Klein-Gordon equation that can be used to propagate the scalar field. One such
form admits a {\it conserved inner product} on the space of complex solutions
to the field equations, and this form gives rise to particle production.  The
amount of particle production depends only on the ``temporal" extent of the
Euclidean region in the spacetime.

This paper attempts to extend the work of Dray {\it et al.}~to the case of a
(massless) spin-half field in a signature changing spacetime.  We shall
show---for $SL(2,C)$ spinors---that although there are inequivalent forms of
the Dirac equation that can be used to propagate a spinor in a signature
changing spacetime, {\it none} of these forms gives rise to a conserved inner
product on the space of solutions to the field equations. We do not know at
present what the implications are of this result. Perhaps the only consequence
is that it is ``incorrect" to use $SL(2,C)$ spinors in signature changing
spacetimes. Or perhaps the standard calculational techniques for analyzing
particle production for spinors have to be modified. Or perhaps it is
impossible to formulate a consistent quantum field theory for spinors in a
signature changing spacetime. We are currently investigating these
issues.\footnote{We should note that other authors, in particular Rubakov
\cite{Rubakov} and Kandrup \cite{Kandrup}, have considered the question of
particle production in spacetimes that have metrics with {\it Euclidean}
signature.  They found that one can extend the standard formulations of quantum
field theory in Lorentzian spacetimes in a way that leads to {\it non-unitary}
Bogolubov transformations \cite{Rubakov}. Their calculations, however, take
advantage of a conserved inner product on the space of complex solutions to the
field equations.}

The organization of the paper is as follows:  In section 2, we will review the
analysis of Dray {\it et al.}~\cite{DM&T} for a (massless) scalar field to
indicate how they obtained their results and to set the stage for the spinor
calculations of the following section.  We will emphasize the importance of
finding a conserved inner product and show that one form of the Klein-Gordon
equation admits such a structure.  We will not explicitly solve the field
equations; we will just quote the result of \cite{DM&T} at the end of section 2
to show that particle production does, indeed, occur.  In section 3, we will
analyze the propagation of a (massless) spin-half field in a signature changing
spacetime.  We first define $SL(2,C)$ spinors in such a spacetime, and then
show that although there are inequivalent forms of the Dirac equation, none of
these forms gives rise to a conserved inner product on the space of solutions
to the field equations. Finally, in section 4, we will make a few remarks about
the general program of trying to formulate a quantum field theory for spinors
in signature changing spacetimes.

For simplicity, we will take our background, signature changing spacetime to be
a real, 4-dimensional manifold $M$ that is topologically $R^4$, has global
coordinates $(\t,\vec x):=(\t,x,y,z)$, and has a metric $g_{\u\v}$ whose
components change {\it discontinuously} at $\t=0$ and $\t=\D$:
$$ds^2=\mp d\t^2+dx^2+dy^2+dz^2.\eqno(1)$$
The minus sign is used for $\t<0$ and $\t>\D$; the plus sign is used for $0\leq
\t\leq\D$. The spacetime is thus spatially homogeneous (in fact, spatially
flat) and has a single Euclidean region of finite ``temporal" extent $\D$.  It
is Minkowskian at early $(\t<0)$ and late $(\t>\D)$ times, so we can define
``in" and ``out" particle states.  Note also that the field equations we have
to solve simplify for such a discontinuous metric. We need only solve the
differential equations in each region separately and then match the solutions
at the boundaries $\t=0$ and $\t=\D$.  Of course, the matching conditions are
determined by the particular form of the full field equation.

%
%

\vskip 1cm
\noindent{\bf 2. Scalar Field}
\vskip .5cm

As mentioned in the introduction, we will begin by considering the propagation
of a (massless) scalar field  in a signature changing spacetime. This
discussion is basically a review of the analysis given by Dray {\it et al.},
although we will make explicit some of the finer details that were implicit in
their work \cite{DM&T}.  In particular, we will emphasize the importance of
finding a form of the Klein-Gordon equation that admits a conserved inner
product and show that one form admits such a structure while another form does
not.\footnote{One can easily extend the following analysis to the case of a
{\it massive} scalar field.  In particular, Eq.~(8) still holds provided
$\w:=|\vec k|$ is replaced by $\w:= (\vec k\cdot\vec k + m^2)^{{1\over 2}}$,
where $m$ denotes the mass of the scalar field.}

The Klein-Gordon equation for a (massless) scalar field $\phi$ in a general
curved spacetime is
$$g^{\u\v}\grad_\u\grad_\v\phi=0,\eqno(2)$$
where $g^{\u\v}$ is the contravariant metric and $\grad_\u$ is the unique
torsion-free, covariant derivative operator compatible with $g_{\u\v}$. In a
Lorentzian spacetime with a non-degenerate metric, Eq.~(2) can be written in
many equivalent forms.  For example, by expanding $\grad_\u$ in terms of a
partial derivative operator $\pd_\u$ and the associated Christoffel symbols, we
find that Eq.~(2) is equivalent to
$${1\over\sqrt{-g}}\pd_\u(\sqrt{-g}\>g^{\u\v}\pd_\v\phi)=0,\eqno(3a)$$
where $g$ denotes the determinant of $g_{\u\v}$.  (Here we have used the
algebraic identity $g^{\v\l}\pd_\u g_{\v\l}=g^{-1}\pd_\u g$.) In addition, we
obtain another form equivalent to Eq.~(2) by replacing $-g$ with $|g|$ in
Eq.~($3a$). Explicitly, we have
$${1\over\sqrt{|g|}}\pd_\u(\sqrt{|g|}\>g^{\u\v}\pd_\v\phi)=0.\eqno(3b)$$
Although there exist other equivalent ways of rewriting Eq.~(2), we will
consider only those given above. In particular, we have chosen to avoid any
form of the Klein-Gordon equation that explicitly involves the Christoffel
symbols.

For signature changing spacetimes, the situation becomes more complicated.
Naively, Eqs.~($3a,b$) are inequivalent due to the distributional derivative of
the absolute value in Eq.~($3b$).  However, one has to be careful when taking
derivatives in signature changing spacetimes, since for our discontinuous
metric (1),  $\t$-derivatives of $g^{\u\v}$ and $\t$-derivatives of $g$ will
involve delta functions. By themselves, the delta functions pose no serious
difficulty, but products of the delta functions with other discontinuous
functions do. These products are in general ambiguous.\footnote{Even the chain
rule fails to hold for something as simple as a product of step functions!
Although one can define the value of the step function $\theta(\t)$ at $\t=0$
so that ${d\over d\t}\theta^2(\t)=2\theta(\t) {d\theta\over d\t}$
($\theta(0):={1\over 2}$ works), it follows that $\d(\t)={d\over
d\t}\theta^3(\t)\not=3\theta^2(\t){d\theta\over d\t}={3\over 4} \d(\t)$.} This
problem can be avoided, however, if we multiply Eqs.~($3a,b$) by $\sqrt{-g}$
and $\sqrt{|g|}$ and interpret the components $\sqrt{-g}\>g^{\u\v}$ and
$\sqrt{|g|}\>g^{\u\v}$ as single, discontinuous densities, rather than products
of the discontinuous densities $\sqrt{-g}$ or $\sqrt{|g|}$ with $g^{\u\v}$.
With such an interpretation,
$$\eqalignnotwo{\pd_\u(\sqrt{-g}\>g^{\u\v}\pd_\v\phi)&=0\quad{\rm and}
&(4a)\cr
\pd_\u(\sqrt{|g|}\>g^{\u\v}\pd_\v\phi)&=0&(4b)\cr}$$
are unambiguous.  For our particular spacetime in the global coordinates
$(\t,\vec x)$, we have
$$\sqrt{-g}\>g^{\u\v}=\sqrt{|g|}\>g^{\u\v}={\rm diag}(-1,1,1,1)\eqno(5a)$$
in the initial $(\t<0)$ and final $(\t>0)$ Lorentzian regions, and
$$\eqalignnotwo{\sqrt{-g}\>g{}^{\u\v}&={\rm diag}(i,i,i,i)\quad{\rm and}
&(5b)\cr
\sqrt{|g|}\>g{}^{\u\v}&={\rm diag}(1,1,1,1)&(5c)\cr}$$
in the Euclidean region $(0\leq\t\leq\D)$. Note that $\sqrt{|g|}\>g{}^{\u\v}$
is real, while $\sqrt{-g}\>g{}^{\u\v}$ is complex. Eqs.~($4a,b$) are
{\it inequivalent} simply because the densities $\sqrt{-g}\>g{}^{\u\v}$ and
$\sqrt{|g|}\>g{}^{\u\v}$ differ.

In addition to the ``mathematical niceties" of Eqs.~($4a,b$), there are, at
least, two other reasons why one might prefer these equations to Eqs.~($3a,b$).
%
%
First, finding a conserved inner product on the space of complex solutions to
the field equations is equivalent to finding a divergence-free vector density
of weight one, $\otw j^{\u}[\phi_1,\phi_2]$, that is anti-linear in the first
argument and linear in the second. Since all such vector densities constructed
from $g_{\u\v}$ and $\phi$ will involve the metric components through the
combination $\sqrt{-g}\>g^{\u\v}$ or $\sqrt{|g|}\>g^{\u\v}$, the most relevant
forms of the Klein-Gordon equation for checking whether $\pd_\u \otw j^\u=0$
are Eqs.~($4a,b$).  Second, since Eqs.~($4a,b$) are total divergences, it is
easy to determine what the associated matching conditions are for our
particular signature changing spacetime. These conditions are obtained by
integrating the equations across the boundaries---that is, by integrating from
$-\e$ to $\e$ and from $(\D-\e)$ to $(\D+\e)$, and then taking the limit as
$\e\rightarrow 0$. From Eqs.~($5a,b,c$), we see that the matching conditions
associated with Eq.~($4a$) involve a factor of $i$, while those associated with
Eq.~($4b$) do not.

Given Eqs.~($4a,b$) as our starting point, the goal is to determine which of
these equations, if any, admits a conserved inner product on the space of
complex solutions to the field equations. Since we are to consider complex
solutions, we need, in addition to Eqs.~($4a,b$), the complex conjugate
equations
$$\eqalignnotwo{\pd_\u((\sqrt{-g}\>g^{\u\v})^*\pd_\v\phi^*)&=0\quad{\rm and}
&(4a^*)\cr
\pd_\u(\sqrt{|g|}\>g^{\u\v}\pd_\v\phi^*)&=0.&(4b^*)\cr}$$
Note that $\phi$ satisfies Eqs.~($4a,b$) if and only if the complex conjugate
field $\phi^*$ satisfies Eqs.~($4a^*,b^*$). Since $\sqrt{-g} \>g^{\u\v}$ is
complex, the field equation ($4a^*$) for $\phi^*$ is not simply Eq.~($4a$) with
$\phi$ replaced by $\phi^*$.

Now we are basically finished.  Using  Eqs.~($4b$) and ($4b^*$), it is easy to
show that the vector density
$$\otw j^\u[\phi_1,\phi_2]:=i\sqrt{|g|}\>g^{\u\v}(\phi_1^*\pd_\v\phi_2
-\phi_2\pd_\v\phi_1^*)\eqno(6a)$$
is divergence-free---i.e., that $\pd_\u\otw j^\u=0$ for complex solutions
$\phi_1$ and $\phi_2$ of Eq.~($4b$). Thus, we obtain a {\it conserved inner
product}, $(\phi_1,\phi_2):=\int_\S\>\otw j^\u[\phi_1,\phi_2]\>d\S_\u$,
by integrating $\otw j^\u$ over any 3-dimensional hypersurface $\S$. On the
other hand, the vector density
$$\otw j^\u[\phi_1,\phi_2]:=i\sqrt{-g}\>g^{\u\v}(\phi_1^*\pd_\v\phi_2
-\phi_2\pd_\v\phi_1^*)\eqno(6b)$$
is not divergence-free for complex solutions $\phi_1$ and $\phi_2$ of
Eq.~($4a$). This is precisely because Eq.~($4a^*$) involves the complex
conjugate $(\sqrt{-g}\>g^{\u\v})^*$.\footnote{For what it's worth, $\pd_\u\otw
j^u=2(\theta(\t)-\theta(\t-\D))\phi_2(\pd_\t^2+\pd_x^2+\pd_y^2
+\pd_z^2)\phi_1^*+2(\d(\t)-\d(\t-\D))\phi_2\pd_\t\phi_1^*$.  We should also
point out that Eq.~($4a$) fails to preserve the reality of the scalar field
$\phi$.  Since $\sqrt{-g}\> g^{\u\v}$ is complex in the Euclidean region
$(0\leq \t\leq\D)$, a real $\phi$ (for $\t<0$) evolves to a complex $\phi$ (for
$\t>\D$).}


Finally, to conclude this section, we will show that Eq.~($4b$) gives rise to
particle production.  We will only quote the final result here---details can be
found in \cite{DM&T}.  It suffices to say that one simply solves Eq.~($4b$) in
each region separately, and then joins these solutions at the boundaries $\t=0$
and $\t=\D$ using the matching conditions implied by Eq.~($4b$).  If we take
for $\t<0$ an initial positive frequency solution
$$\phi_{{\rm in}}(\vec x,\t)=e^{i(\vec k\cdot\vec x- \w\t)}\quad{\rm where}
\quad \w:=|\vec k|,\eqno(7)$$
we find for $\t>\D$,
$$\phi_{{\rm out}}(\vec x,\t)=e^{i\vec k\cdot\vec x}[\cosh(\w\D)e^{-i\w(\t-\D)}
+i\sinh(\w\D)e^{i\w(\t-\D)}].\eqno(8)$$
Since the intial positive frequency solution evolves to a final solution
containing a mixture of positive and negative frequency parts, particle
production does, indeed, occur. The amount of particle production for a given
mode is proportional to $\sinh^2(\w\D)$ and depends only on the ``temporal"
extent of the Euclidean region in the spacetime.\footnote{If the boundaries
$\t=0$ and $\t=\D$ are given instead by $\t=\t_i$ and $\t=\t_f$ where
$\t_f-\t_i=\D$, Eq.~(8) would still hold provided the phase factor
$e^{i\w(\t-\D)}$ multiplying the $\sinh(\w\D)$ term is changed to $e^{i\w
(\t-(\t_i+\t_f))}$.} This is the main result of Dray {\it et al.}~\cite{DM&T}.

\vskip 1cm
\noindent{\bf 3. Spinor Field}
\vskip .5cm

To begin our discussion of (massless) spin-half fields, we must first define
what we mean by a spinor field in a signature changing spacetime.  In this
paper, we choose to fix the underlying spin group to be $SL(2,C)$ throughout
the spacetime. We shall see that this definition of $SL(2,C)$ spinors for
signature changing spacetimes requires the soldering form to be {\it complex}
in the Euclidean region.

Recall that the standard definition of $SL(2,C)$ spinors for a 4-dimensional
Lorentzian spacetime \cite{Wald} starts with a 2-dimensional complex vector
space $W$ that is equipped with a non-degenerate, anti-symmetric tensor
$\e_{AB}$. \ $\e_{AB}$ and its inverse $\e^{AB}$ are normalized so that
$\e_{AB}\e^{AB}=2$, and they are used to lower and raise spinor indices
according to the conventions $\l^A\e_{AB}=:\l_B$ and $\e^{AB}\u_B=:\u^A$. In
addition to $W$, we have the complex conjugate vector space $\ovr W$ that is in
1-1, anti-linear correspondence with $W$.\footnote{If we let $W^*$ denote the
dual space of $W$ (i.e., the complex vector space of linear mappings from $W$
to $C$), then $\ovr W$ is the complex vector space of {\it anti}-linear
mappings from $W^*$ to $C$.}  For $\l^A\in W$, the corresponding element in
$\ovr W$ will be denoted by $\ovr\l{}^{A'}$. In particular, the complex
conjugate anti-symmetric tensor is $\ovr\e_{A'B'}$. Now it is fairly easy to
show that the subspace $V\subset W\tprod\ovr W$ consisting of elements
$\a^{AA'}$ satisfying $\ovr\a{}^{AA'}=-\a^{AA'}$ is a 4-dimensional {\it real}
vector space. Moreover, it is equipped with a metric $\n_{AA'BB'}:=\e_{AB}
\ovr\e_{A'B'}$ of Lorentzian signature $(-+++)$.  Thus, if the spacetime metric
$g_{\u\v}$ is also of Lorentzian signature, there is an isomorphism
$\s^\u{}_{AA'}(p)$ from $V$ to the tangent space to $M$ at $p$ satisfying
$\s^\u{}_{AA'}\s^\v{}_{BB'}\n^{AA'BB'}=g^{\u\v}$. \ $\s^\u{}_{AA'}$ is often
called a {\it soldering form} since it ``solders" elements $\a^{AA'}\in V$ to
tangent vectors $\a^\u:=\s^\u{}_{AA'}\a^{AA'}$. If $\s^\u{}_{AA'}$ exists
globally on $M$, then $M$ is said to admit an $SL(2,C)$ spinor structure. \
$\l^A$ is then called an $SL(2,C)$ spinor. Note that since tangent vectors are
real, \ $\ovr\s{}^\u{}_{AA'}= -\s^\u{}_{AA'}$.

If the spacetime metric $g_{\u\v}$ has Euclidean signature $(++++)$, then there
no longer exists an isomorphism $\s^\u{}_{AA'}(p)$ from $V$ to the
4-dimensional (real) tangent space to $M$ at $p$ satisfying
$\s^\u{}_{AA'}\s^\v{}_{BB'}\n^{AA'BB'}=g^{\u\v}$. If we drop the requirement
that $\s^\u{}_{AA'}$ be an isomorphism, but still demand that
$\s^\u{}_{AA'}\s^\v{}_{BB'}\n^{AA'BB'}=g^{\u\v}$, we find that $\s^\u{}_{AA'}$
maps $V$ to the {\it complexified} tangent space to $M$ at $p$. For such a
$\s^\u{}_{AA'}$, the tangent vector $\a^\u:=\s^\u{}_{AA'}\a^{AA'}$ is, in
general, complex. Because of this, $\ovr\s{}^\u{}_{AA'}$ is no longer
proportional to $\s^\u{}_{AA'}$.

Note that the above construction in terms of soldering forms can be expressed
equally well in terms of tetrads.  One simply chooses, once and for all, four
orthonormal basis elements $\t^{\und a}{}_{AA'}\in V$, where $\und a=0,1,2,3$.
These satisfy $\ovr\t{}^{\und a}{}_{AA'}= -\t^{\und a}{}_{AA'}$ and $\t^{\und
a}{}_{AA'}\t^{\und b}{}_{BB'}\n^{AA'BB'} =\n^{\und a\und b}:={\rm
diag}(-1,1,1,1)$. Then one defines $e_{\und a}^\u$ by
$\s^\u{}_{AA'}(p)=:e_{\und a}^\u(p)\t^{\und a}{}_{AA'}$.  Since $\t^{\und a}
{}_{AA'}$ maps $\n^{AA'BB'}$ to $\n^{\und a\und b}$, we have $g^{\u\v}= e_{\und
a}^\u e_{\und b}^\v\n^{\und a\und b}$, i.e., $e_{\und a}^\u$ is a tetrad.  For
Lorentzian spacetimes, tetrads can be chosen to be real. But in the Euclidean
region of a signature changing spacetime, tetrads are necessarily complex (in
any gauge). This is because we have chosen to fix the signature of the
Minkowski metric $\n^{\und a\und b}$ to be $(-+++)$ throughout the spacetime.
(See, also, pp.21-4 of ref.\cite{P&R1} and pp.459-60 of ref.\cite{P&R2}.)

We are finally ready to write down the Dirac equation for a (massless)
spin-half field $\psi^A$. In a general curved spacetime, the Dirac equation is
$$\s^\u{}_{AA'}\grad_\u\psi^A=0,\eqno(9)$$
where $\s^\u{}_{AA'}$ is an $SL(2,C)$ soldering form and $\grad_\u$ is the
unique torsion-free, spin-covariant derivative operator compatible with
$\s^\u{}_{AA'}$.  In a Lorentzian spacetime with a non-degenerate metric,
Eq.~(9) can be written in many equivalent forms.  For example, by expanding
$\grad_\u$ in terms of a partial derivative operator $\pd_\u$ and the
associated spin-connections $\w_\u{}^{AB}$ and $\w_\u{}^{A'B'}$, we find that
Eq.~(9) is equivalent to
$${1\over\sqrt{-g}}\pd_\u(\sqrt{-g}\>\s^\u{}_{AA'}\psi^A)+\w_{\u A'}{}^{B'}
\s^\u{}_{AB'}\psi^A=0.\eqno(10a)$$
In addition, we obtain another form equivalent to Eq.~(9) by replacing $-g$
with $|g|$ in Eq.~($10a$). Explicitly, we have
$${1\over\sqrt{|g|}}\pd_\u(\sqrt{|g|}\>\s^\u{}_{AA'}\psi^A)+\w_{\u A'}{}^{B'}
\s^\u{}_{AB'}\psi^A=0.\eqno(10b)$$
Although there exist other equivalent ways of rewriting Eq.~(9), we will
consider only those given above.\footnote{One might ask whether I am leaving
out any ``important" forms of the Dirac equation by considering only
Eqs.~($10a,b$). I believe the answer is no, since, as I will argue shortly,
only certain forms of the Dirac equation are unambiguous for our particular
signature changing spacetime with discontinuous metric (1). These turn out to
be  Eqs.~($11a,b$) with $\w_\u{}^{A'B'}=0$. All other forms of the Dirac
equation (at least all those that I could think of that involve $SL(2,C)$
spinors) are either ambiguous or reduce to one of these.}

For signature changing spacetimes, the situation again becomes more
complicated. For the same reasons we gave in section 2 for the scalar field, we
consider, instead of Eqs.~($10a,b$), the equations
$$\eqalignnotwo{\pd_\u(\sqrt{-g}\>\s^\u{}_{AA'}\psi^A)+\w_{\u A'}{}^{B'}
\sqrt{-g}\>\s^\u{}_{AB'}\psi^A&=0\quad{\rm and}&(11a)\cr
\pd_\u(\sqrt{|g|}\>\s^\u{}_{AA'}\psi^A)+\w_{\u A'}{}^{B'}\sqrt{|g|}\>
\s^\u{}_{AB'}\psi^A&=0,&(11b)\cr}$$
where the components $\sqrt{-g}\>\s^\u{}_{AA'}$ and $\sqrt{|g|}\>\s^\u
{}_{AA'}$ are to be thought of as single, discontinuous densities rather than
products of the discontinuous densities $\sqrt{-g}$ or $\sqrt{|g|}$ with
$\s^\u{}_{AA'}$. But now there is an additional complication.  Eqs.~($11a,b$)
are well-defined only if the terms $\w_\u{}^{A'B'}
\sqrt{-g}\>\s^\u{}_{AB'}\psi^A$ and $\w_\u{}^{A'B'}\sqrt{|g|}\>\s^\u{}_{AB'}
\psi^A$ do not involve ambiguous products of delta functions with other
discontinuous functions.  Equivalently,  Eqs.~($11a,b$) are well-defined only
if there exists a gauge---i.e., a choice of tetrad $e_{\und a}^\u$ or soldering
form $\s^\u{}_{AA'}$---in which the spin-connections $\w_\u{}^{AB}$ and
$\w_\u{}^{A'B'}$ are determined uniquely and do not involve any delta
functions. As we will now show, for our particular signature changing spacetime
with discontinuous metric (1), there always exists such a gauge.

To see this, it is convenient to work with a tetrad $e_{\und a}^\u$ and the
corresponding Ricci rotation coefficients $\w_\u{}^{\und a\und b}$. In terms of
the spin-connections $\w_\u{}^{AB}$ and $\w_\u{}^{A'B'}$, the Ricci rotation
coefficients are given by $\w_\u{}^{\und a\und b}= \t^{\und a}{}_{AA'}\t^{\und
b}{}_{BB'}(\w_\u{}^{AB}\ovr\e{}^{A'B'}+ \w_\u{}^{A'B'}\e{}^{AB})$. They are
determined uniquely by $e_{\und a}^\u$ and their inverses $e_\u^{\und a}$ via
the differential equation $de^{\und a}+\w^{\und a}{}_{\und b}\wedge e^{\und
b}=0$. For the discontinuous metric (1), we can choose
$$e_{\und a}^\u=(\pd_\t,\pd_x,\pd_y,\pd_z)\eqno(12a)$$
in the initial and final Lorentzian regions and
$$e_{\und a}^\u=(i\pd_\t,\pd_x,\pd_y,\pd_z)\eqno(12b)$$
in the Euclidean region, so that $e_\u^{\und a}=(d\t,dx,dy,dz)$ and $e_\u^{\und
a}=(-id\t,dx,dy,dz)$. Then $de^{\und a}=0$.\footnote{Even though $e^{\und 0}$
changes discontinuously at $\t=0$ and $\t=\D$, we have $d e^{\und
0}=(-i-1)(\d(\t)-\d(\t-\D))d\t\wedge d\t=0$.} Since $e^{\und b}$ is invertible,
$\w_\u{}^{\und a\und b}=0$ as well. Hence, the spin-connections $\w_\u{}^{AB}$
and $\w_\u{}^{A'B'}$ are zero, so Eqs.~($11a,b$) are well-defined.


Thus, in the gauge specified by (12), the two forms of the Dirac equation, Eqs.
($11a,b$), can be written as
$$\eqalignnotwo{\pd_\u(\sqrt{-g}\>e_{\und a}^\u\t^{\und a}{}_{AA'}\psi^A)&=0
\quad{\rm and}&(13a)\cr
\pd_\u(\sqrt{|g|}\>e_{\und a}^\u\t^{\und a}{}_{AA'}\psi^A)&=0.&(13b)\cr}$$
As before, the components $\sqrt{-g}\>e_{\und a}^\u$ and $\sqrt{|g|}\> e_{\und
a}^\u$ are to be thought of as single, discontinuous densities rather than
products of the discontinuous densities $\sqrt{-g}$ or $\sqrt{|g|}$ with
$e_{\und a}^\u$. For our particular spacetime and the above choice of gauge, we
have
$$\sqrt{-g}\>e_{\und a}^\u=\sqrt{|g|}\>e_{\und a}^\u=(\pd_\t,\pd_x,\pd_y,
\pd_z)\eqno(14a)$$
in the initial and final Lorentzian regions, and
$$\eqalignnotwo{\sqrt{-g}\>e_{\und a}^\u&=(-\pd_\t,i\pd_x,i\pd_y,i\pd_z)
\quad{\rm and}&(14b)\cr
\sqrt{|g|}\>e_{\und a}^\u&=(i\pd_\t,\pd_x,\pd_y,\pd_z)&(14c)\cr}$$
in the Euclidean region. Note that both $\sqrt{-g}\>e_{\und a}^\u$ and
$\sqrt{|g|}\>e_{\und a}^\u$ are complex.   Eqs.~($13a,b$) are {\it
inequivalent} simply because the densities $\sqrt{-g}\>e_{\und a}^\u$ and
$\sqrt{|g|}\>e_{\und a}^\u$ differ.

Given  Eqs.~($13a,b$) as our starting point, the goal is to determine if either
of these equations admits a conserved inner product on the space of solutions
to the field equations. Just as for the scalar field, we need, in addition to
Eqs.~($13a,b$), the complex conjugate equations
$$\eqalignnotwo{\pd_\u((\sqrt{-g}\>e_{\und a}^\u)^*\t^{\und a}{}_{AA'}\ovr
\psi{}^{A'})&=0\quad{\rm and}&(13a^*)\cr
\pd_\u((\sqrt{|g|}\>e_{\und a}^\u)^*\t^{\und a}{}_{AA'}\ovr\psi{}^{A'})&=0.
&(13b^*)\cr}$$
Since both $\sqrt{-g}\>e_{\und a}^\u$ and $\sqrt{|g|}\>e_{\und a}^\u$ are
complex, the field equations for $\ovr\psi{}^{A'}$ are not simply
Eqs.~($13a,b$) with $\psi^A$ replaced by $\ovr\psi{}^{A'}$.

But now note that neither form of the Dirac equation, Eqs.~($13a,b$), gives
rise to a conserved inner product. The vector density
$$\otw j^\u[\psi_1,\psi_2]:=i\sqrt{-g}\>e_{\und a}^\u\t^{\und a}{}_{AA'}
\psi_2^A\ovr\psi{}_1^{A'}\eqno(15a)$$
is not divergence-free for solutions $\psi{}_1^A$ and $\psi_2^A$ of
Eq.~($13a$). Similarly, the vector density
$$\otw j^\u[\psi_1,\psi_2]:=i\sqrt{|g|}\>e_{\und a}^\u\t^{\und a}{}_{AA'}
\psi_2^A\ovr\psi{}_1^{A'}\eqno(15b)$$
is not divergence-free for solutions $\psi{}_1^A$ and $\psi_2^A$ of
Eq.~($13b$). This is precisely because  Eqs.~($13a^*$) and ($13b^*$) involve
the complex conjugates $(\sqrt{-g}\>e_{\und a}^\u)^*$ and $(\sqrt{|g|}\>e_{\und
a}^\u)^*$.

\vskip 1cm
\noindent{\bf 4. Conclusion}
\vskip .5cm

In light of the above results, we see, at the very least, that formulating
quantum field theory in signature changing spacetimes is not going to be an
easy task. Since a (massless) spin-half field in a signature changing spacetime
does not admit a conserved inner product on the space of solutions to the field
equations, we are forced to step back and rethink what structures are essential
for setting up a consistent mathematical theory.  As mentioned in the
introduction, the lack of a conserved inner product might be due to an
``incorrect" definition of spinors in signature changing spacetimes. Perhaps
the underlying spin group should be changed from $SL(2,C)$ to $SU(2)\times
SU(2)$ in the Euclidean region. Rather than extend the definition of $SL(2,C)$
spinors to signature changing spacetimes as we did in this paper, one could
possibly switch to a non-standard, 8-dimensional (real) representation of
spinors in the Lorentzian regions that can be mapped onto an 8-dimensional
(real) representation of spinors in the Euclidean region.\footnote{I am
indebted to an anonymous referee who pointed this out. For definitions of
spinors in arbitrary dimensions and for arbitrary signatures, see
\cite{P&R2,spinors,Mehta}.} Or perhaps the definition of spinors we gave is
``correct," but the standard calculational techniques for analyzing particle
production have to be modified for signature changing spacetimes. Neither of
these two cases represents a serious problem. However, it might turn out that
the absence of a conserved inner product means that one simply cannot do
quantum field theory for spinors in a signature changing spacetime. We do not
know the answer to these questions at present.

Finally, there is a problem even for the case of a (massless) scalar field with
a conserved inner product.  Since Eq.~(8) shows that the amount of particle
production for a given mode is proportional to $\sinh^2(\w\D)$,  the total
energy of the produced particles diverges. Thus, the back reaction of the
produced particles on the signature changing spacetime cannot be ignored.
Further investigation of the scalar field case would have to address this
issue.

\vspace{1cm}

\noindent ACKNOWLEDGEMENTS

\vspace{.3cm}

I would like to thank Arlen Anderson, Bei-Lok Hu, and Kay Pirk for many helpful
discussions. As usual, Ted Jacobson provided numerous suggestions that greatly
improved the quality of this paper. I would also like to thank Tevian Dray for
clarifying certain remarks that I made regarding the scalar field, and an
anonymous referee for constructive criticism regarding the spin-half field.
This work was supported in part by NSF grant PHY91-12240.

\end{document}